\documentclass[runningheads]{llncs}
\usepackage[T1]{fontenc}
\usepackage{graphicx}
\usepackage[utf8]{inputenc}
\usepackage{authblk,amssymb}
\usepackage{setspace}
\usepackage[margin=1.25in]{geometry}
\usepackage{graphicx}
\graphicspath{ {./figures/} }
\usepackage{subcaption}
\usepackage{amsmath}
\usepackage{lineno}
\usepackage{pgfplots}
\usepackage{filecontents}
\usepackage[misc]{ifsym}
\usepackage{soul}
\def\wt{{\rm wt}}
\def\ord{{\rm ord}}
\def\Fix{{\rm Fix}}
\def\Z{\mathbb{Z}}

\newcommand{\F}{\mathbb{F}}
\usepackage{hyperref}
\hypersetup{colorlinks=true,
      linkcolor=blue}
\begin{document}
\title{Quantum CSS Duadic and Triadic Codes: New Insights and Properties}

\author{Reza Dastbasteh \Letter \orcidID{0009-0009-2104-7485} \and
Olatz Sanz Larrarte\orcidID{0009-0003-5744-8471} \and
Josu Etxezarreta Martinez\orcidID{0000-0001-7058-8426} \and
Antonio deMarti iOlius\orcidID{0000-0003-0602-5535} \and
Javier Oliva del Moral\orcidID{0009-0002-9309-9347} \and
Pedro Crespo Bofill\orcidID{0000-0001-8259-3353}
}

\authorrunning{R. Dastbasteh, O. Sanz, J. Etxezarreta, A. deMarti, J. Oliva, P. Crespo}

\institute{Tecnun\ - University of Navarra, Donostia-San Sebastian, Spain,\\
\email{\{rdastbasteh,osanzl,jetxezarreta,ademartio,jolivam,pcrespo\}@unav.es}}

\maketitle       
\begin{abstract}
In this study, we investigate the construction of quantum CSS duadic codes with dimensions greater than one. 
We introduce a method for extending smaller splittings of quantum duadic codes to create larger, potentially degenerate quantum duadic codes. 
Furthermore, we present a technique for computing or bounding the minimum distances of quantum codes constructed through this approach. 
Additionally, we introduce quantum CSS triadic codes, a family of quantum codes with a rate of at least $\frac{1}{3}$. 

\keywords{quantum code \and duadic code \and triadic code \and degenerate code \and minimum distance.}
\end{abstract}

\newcommand{\Cl}{{\rm Cliff}}
\newcommand{\Bi}{{\rm bitwise}}

\section{Introduction}
%%%%%%%%%%%%%%%%%%%%%%%%%%%%%%%%%%%%%%%%%%%%%%
Quantum error-correcting codes, also known as quantum codes, serve to protect quantum information from the detrimental effects of noise, such as decoherence, during its transmission, storage or processing.
Analogous to classical error-correcting codes, quantum codes play an important role in ensuring the fidelity of quantum information. Among the various techniques employed for constructing quantum codes, the stabilizer construction stands out as one of the most widely used \cite{Calderbank,Gottesman}. Indeed, this approach has been instrumental in constructing numerous families of quantum codes. 
A notable instance of the stabilizer construction is the class of Calderbank-Steane-Shor (CSS) codes. 
The binary CSS construction offers a straightforward method for generating binary quantum codes by utilizing a binary code and one of its subcodes.

In \cite{aly2006}, it was demonstrated that the CSS construction is well-suited for binary duadic codes, giving rise to an infinite class of quantum duadic codes. 
A notable property of quantum duadic codes is the presence of an infinite subclass of degenerate codes. 
In general, a quantum stabilizer code is called degenerate if its defining stabilizer group contains a non-trivial error with weight smaller than the minimum distance that, thus, acts trivially on the code \cite{Calderbank}. It is important to note that error patterns resulting in non-trivial syndromes necessitate proactive correction of the physical qubits in the quantum code. 
This correction process may introduce additional errors into the system, stemming from the noise introduced by quantum gates. 
Consequently, the construction of highly degenerate codes can potentially reduce the requirement for active error correction. Moreover, codes with degenerate low-weight errors have the potential to perform better since many of the high probability errors do not require to be corrected \cite{DuadicConsta,degenboost}.

One of the primary challenges in constructing degenerate codes lies in determining the true minimum distance or establishing bounds on the minimum distances of these codes. In the case of the CSS duadic quantum codes presented in \cite{aly2006}, all the degenerate codes have a composite length, dimension one, and satisfy a square root minimum distance lower bound.

In this work, inspired by the methodology outlined in \cite{aly2006,DuadicConsta}, we explore the construction of binary quantum duadic codes with larger dimensions using the CSS construction. This is enabled by allowing the multipliers of duadic codes to fix more than one element.
Furthermore, we investigate the existence of such codes and propose a method for extending two smaller length splittings to a larger length splitting. 
This approach results in a class of degenerate binary quantum codes.
Another advantage of the extended splitting method is that it allows us to compute or bound the true minimum distance of quantum codes constructed in this way.
Moreover, we introduce quantum triadic codes, which can be used to build binary quantum codes with rates greater than or equal to $\frac{1}{3}$. Constructing codes with high rates is specially relevant in order to reduce resource consumption. While these constructions might not be the most suitable for integration in state-of-the-art quantum processors as a result of connectivity constraints, they can be useful for future processors or quantum communications \cite{qcomms}.
Finally, we present the parameters for several quantum duadic and triadic codes constructed using our approach. 

This paper is structured as follows. In Section \ref{S:Priliminary}, we give an overview of the structure and properties of cyclic, duadic, and quantum stabilizer codes. Section \ref{S:Existence} discusses the existence of quantum CSS duadic codes with dimensions larger than one, along with their construction.
In Section \ref{S:Extended splitting}, we explore the properties of quantum CSS duadic codes constructed from extended splittings, their degenerate subclass, and methods for computing or bounding their minimum distances. Section \ref{S:Triadic} introduces quantum triadic codes and their properties.
Finally, Section \ref{S:Conc} summarizes our results.

%%%%%%%%%%%%%%%%%%%%%%%%%%%%%%%%%%%%%%%%%%%%%%%%%%%%%%%%%%%%%%%%%%
\section{Preliminaries}\label{S:Priliminary}
%%%%%%%%%%%%%%%%%%%%%%%%%%%%%%%%%%%%%%%%%%%%%%%%%%%%%%%%%%%%%%%%%%%

Let $\F_2$ be the binary field. Throughout this work, whenever we discuss cyclic or duadic codes, $n$ is a positive odd integer.

\subsection{Cyclic and Duadic Codes}
Duadic codes encompass a variety of well-known codes such as quadratic residue codes, Golay codes, and numerous Reed-Muller and Reed-Solomon codes, all of which have rich algebraic structure and can be used to construct good classical and quantum codes \cite{DuadicConsta,Pless2,Smid,RezaDuadic,guenda2009}, and \cite[Section 6.5]{Huffman}. Given that  they are cyclic, we will commence with a brief overview of cyclic codes.

A linear code $C\subseteq \F_2^n$ is called  \textit{cyclic} if for every $c=(c_0,c_1,\ldots,c_{n-1})\in C$, the vector $(c_{n-1},c_0,\ldots,c_{n-2})$ obtained by a cyclic shift of the coordinates of $c$ is also in $C$. 
It is well known that there exists a one-to-one correspondence between binary cyclic codes of length $n$ and ideals of the ring $\F_2 [x]/\langle x^n-1\rangle $, for example see \cite[Section 4.2]{Huffman}. Thus each cyclic code can be uniquely represented by a monic polynomial $g(x)$, where $g(x)$ is the minimal degree generator of the corresponding ideal. The polynomial $g(x)$ is called the {\em generator polynomial} of such cyclic code. Let $\alpha$ be a
primitive $n$-th root of unity in a finite field extension of $\F_2$. Alternatively, such cyclic code can be represented by its unique {\em defining set}
\begin{equation}\label{E:Defining}
\{t: 0\leq t \le n-1 \ \text{and}\ g(\alpha^t)=0 \}.
\end{equation} 
As we will see, many of the characteristics of cyclic codes are easily determined by the notion of a defining set. For each $s\in \mathbb{Z}/n\Z$, the \textit{$2$-cyclotomic coset} modulo $n$ containing $s$ is defined as 
\begin{equation}
Z(s)=\{(s2^j) \mod n: 0\le j \le m-1\},
\end{equation}
where $m$ is the smallest positive integer such that $s2^m \equiv s \pmod{n}$. All different  $2$-cyclotomic cosets partition $\mathbb{Z}/n\Z$, and the defining set
of a linear cyclic code is a union of cyclotomic cosets.
A {\em multiplier} $\mu_b$ on $\Z/n\Z$ is defined by $\mu_b(x)= (bx) \mod{n}$, for some integer $b$ such that $\gcd(n,b)=1$. 
Note also that if $A \subseteq \Z/n\Z$, then $\mu_{b}(A)=bA=\{(ba) \mod n: a \in A\}$.
Multipliers act as special permutations on $\mathbb{Z}/n\mathbb{Z}$ and they are essential to define duadic and related codes. 
If $C$ is a cyclic code with the defining set $A$, then $\mu_b(C)$, i.e., the code obtained by permuting the coordinates of $C$ by applying $\mu_b$, is a cyclic code with the defining set $\mu_{b^{-1}}(A)=b^{-1}A$.

Recall that the Euclidean inner product of $u=(u_0,u_1,\ldots,u_{n-1})$ and $v=(v_0,v_1,\ldots,v_{n-1})$ is defined by 
\[u\cdot v=\sum_{i=0}^{n-1}u_iv_i \in \F_2.\]

The (Euclidean) dual of a binary cyclic code $C$ of length $n$ is defined by 
\[C^{\bot}=\{u\in\F_2^n: u\cdot c=0 \ \text{for each} \ c \in C \}.\]
If a cyclic code $C$ has defining set $A$, then the set 
\[  (\mathbb{Z}/n\Z) \setminus \mu_{-1}(A)=(\mathbb{Z}/n\Z) \setminus (-A)=(\mathbb{Z}/n\Z) \setminus \{(-a) \mod n: a \in A\}\]
is the defining set of $C^\bot$ \cite[Theorem 4.4.9 (iv)]{Huffman} (in \cite{Huffman} $\Z/n\Z$ and $A$ are denoted by $\mathcal{N}$ and $T$, respectively).
One of our main topics of interest in this paper are dual-containing cyclic codes, i.e., when $C^\bot \subseteq C$.  
The following theorem determines when a binary cyclic code is dual-containing. 

\begin{theorem}\label{T:Euclidean}\label{T:BCH}
Let $C\subseteq \F_2^n$ be a binary cyclic code with the defining set $A$. Then $C^{\bot}\subseteq C$ if and only if $A \cap -A=\emptyset.$
\end{theorem}

\begin{proof}
The proof follows from \cite[Theorem 4.4.11]{Huffman}, and here we only give a short proof for clarification. 
The code $C^\bot$ has the defining set $(\Z/n\Z) \setminus (-A)$. Then
we have $C^{\bot}\subseteq C$ if and only if $A \subseteq  \big( (\Z/n\Z) \setminus (-A) \big)$, which happens if and only if $A \cap -A=\emptyset$.
\hfill $\square$ \end{proof}

Recall that the \emph{(Hamming) weight} of a binary vector $v$ is defined as the number of its non-zero coordinates, denoted by $\wt(v)$. The \emph{minimum (Hamming) distance} of a binary code $C$ is defined as
\[ d(C) = \min\{\wt(c): 0 \neq c \in C \}. \]
Another feature of cyclic codes is that their minimum distance is lower bounded by the Bose-Chaudhuri-Hocquenghem (BCH) bound. 
%The next theorem describes the BCH bound for binary cyclic codes. 
For integers $a\le b$, we denote the consecutive set $\{a,a+1,\ldots,b\}$ by $[a,b]$.

\begin{theorem}\cite{BCH1,BCH2}
Let $C$ be a binary cyclic code of length $n$
with the defining set $A$. If $[a,b] \subseteq A$, then $d(C) \ge (b-a)+2$.    
\end{theorem}

Duadic codes are an important subclass of cyclic codes with rich algebraic and combinatorial properties \cite{Leon-Pless,Pless,Pless3}. Quadratic residue (QR) codes are also a special case of duadic codes, and many examples of linear codes with optimal parameters are constructed from duadic codes. 
% Binary duadic codes were first introduced by Leon et al. \cite{Leon-Pless}, and were later generalized to larger fields by Pless \cite{Pless,Pless3}.
For a more extensive discussion of duadic code see \cite[Chapter 6]{Huffman} and \cite[Section 2.7]{Encyclopedia}.  
In what follows, we provide a more general definition of binary duadic codes based on the concept introduced in Definition 2.6 of \cite{DuadicConsta}, which allows multipliers to fix more than one element.

\begin{definition} \label{splitting def}
Let $X,S_1,S_2 \subseteq \Z/n\Z$ be unions of $2$-cyclotomic cosets such that 
\begin{enumerate}
\item $X\cup S_1 \cup S_2 =\Z/n\Z$,
\item $S_1,S_2$, and $X$ are non-empty and disjoint,
\item there exists a multiplier ${\mu}_b$ such that ${\mu}_b(S_1)=S_2$, ${\mu}_b(S_2)=S_1$, and $\mu_b(Z(s))=Z(s)$ for each $s\in X$.
\end{enumerate}
Then the triple $(X,S_1,S_2)$ is called a {\em splitting} ({\em 2-splitting}) of $\Z/n\Z$ that is given by $\mu_b$. 
\end{definition}

Let $X \subseteq \Z/n\Z$. Recall that $\alpha$ is a primitive $n$-th root of unity in a finite field extension of $\F_2$.
 A vector $(c_0,c_1,\ldots,c_{n-1})\in \F_2^n$ is called {\em even-like with respect to $X$} provided that 
$c(\alpha^s)=0 \ \text{for each} \ s \in X,$
where
$c(x)=\displaystyle\sum_{i=0}^{n-1}c_ix^i \in \F_2[x]$. 
We call a cyclic code \emph{even-like with respect to $X$} if it has only even-like codewords with respect to $X$. Otherwise we call it an {\em odd-like code with respect to $X$}. 
In particular, a linear cyclic code with the defining set $A$ is even-like with respect to any $X\subseteq A$.
Whenever the set $X$ is clear from the context, we simply call the mentioned codes even-like and odd-like. 

\begin{definition}\label{D:duadic}
Let $n$ be a positive odd integer and $(X,S_1,S_2)$ be a splitting of $\Z / n\Z$. Then binary cyclic codes of length $n$ with the defining sets $S_1$ and $S_2$ (respectively $S_1 \cup X$ and $S_2 \cup X$) are called odd-like (respectively even-like) {\em duadic} codes.     
\end{definition} 
Previously, the focus was primarily on duadic codes with a splitting that satisfies $X=\{0\}$, and less attention was given to duadic codes with $|X|>1$. However, as we will explore later, the latter codes can also be used to construct families of classical and quantum codes with similar properties.

Let $C$ and $D$ be odd-like and even-like binary duadic codes of length $n$ with the defining sets $S_1$ and $S_1\cup X$, respectively.  
The {\em minimum odd-like weight} of $C$ (with respect to $X$) is defined by 
$$d_o(C)=\min\{\wt(c):c \in C \setminus D\}.$$
The next theorem highlights some properties of binary duadic codes in this general setting.

\begin{theorem}\label{T:Duadic-properties}
Let $n$ be a positive odd integer and $(X,S_1,S_2)$ be a splitting of $\Z/n\Z$ given by $\mu_{b}$. Let $C_1$ and $C_2$ (respectively $D_1$ and $D_2$) be odd-like (respectively even-like) duadic codes of length $n$ with the defining sets $S_1$ and $S_2$ (respectively $S_1 \cup X$ and $S_2 \cup X$). Then the following hold for each $1 \le i \le 2$: 
\begin{enumerate}
    \item $\mu_{b}(C_1)=C_2$ and $\mu_{b}(D_1)=D_2$ (the pairs of duadic codes are permutation equivalent). 
    \item $\dim(C_i)=\frac{n+|X|}{2}$ and $\dim(D_i)=\frac{n-|X|}{2}$.
    \item $C_1 +C_2=\F_2^n$
 and $D_1 \cap D_2=\{(0,0,\ldots,0)\}$.
    \item $\mu_{-1}(S_1)=S_2$ (or equivalently $\mu_{-1}(S_2)=S_1$) if and only if $D_1^{\bot}=C_1$ and $D_2^{\bot}=C_2$. Also,
    $\mu_{-1}(S_i)=S_i$  if and only if $D_1^{\bot}=C_2$ and $D_2^{\bot}=C_1$. Moreover, the codes $(D_1 +D_2)^{\bot}$ and $C_1 \cap C_2$ are permutation equivalent.
    \item $d_o(C_i)^2 \ge d(C_1 \cap C_2)$ (the code $C_1 \cap C_2$ has the defining set $(\Z/n\Z) \setminus (X)$). 
\end{enumerate}
\end{theorem}

\begin{proof}
The proof is similar to that of \cite[Theorem 3]{Leon-Pless},\cite[Theorem 4.1]{DuadicConsta}, and \cite[Theorems 6.4.2 and 6.4.3]{Huffman}, and we only give a remark about part (4). The code $(D_1 +D_2)^{\bot}$ and $C_1 \cap C_2$ have the defining sets $A=(\Z/n\Z) \setminus (-X)$ and $-A=(\Z/n\Z) \setminus (X)$. 
Thus, they are permutation equivalent (see, for example, \cite[Theorem 1.6.4]{RezaThesis}).
\hfill $\square$ \end{proof}

In Section \ref{S:Existence}, we investigate the existence of binary duadic codes when $|X|>1$.

%%%%%%%%%%%%%%%%%%%%%%%%%%%%%%%%%%%%%%%%%%%%%%%%%%%%%%%%%%%%%%%%%%
\subsection{Binary Quantum Codes}
%%%%%%%%%%%%%%%%%%%%%%%%%%%%%%%%%%%%%%%%%%%%%%%%%%%

In this section, $\mathcal{H}=\mathbb{C}$ is the complex Hilbert space
with the inner product
$\langle a,b \rangle=\displaystyle\sum_{i=1}^{n} \overline{a_i} b_i,$
where $a,b \in \mathcal{H}^n$ and $\overline{a_i}$ is the complex conjugate of $a$. The basic unit of quantum information, i.e., the qubit, lies in the two-dimensional complex Hilbert space, $\mathcal{H}^2$.
The set of one qubit Pauli matrices are defined as 
\begin{center}
$I=\begin{bmatrix}1&0\\0&1\end{bmatrix}$, $X=\begin{bmatrix}0&1\\1&0\end{bmatrix}$, 
$Y=\begin{bmatrix}0&-i\\i&0\end{bmatrix}$, and $Z=\begin{bmatrix}1&0\\0&-1\end{bmatrix}$.
\end{center}
They form a basis for all $2\times 2$ complex matrices. Hence an arbitrary qubit operator ($2\times 2$ unitary matrix over $\mathcal H$) is a linear combinations of the mentioned matrices. 
The set 
\[P_1=\{\pm 1,\pm i\} \times \{ I,  X,  Y,  Z\}\]
 is a multiplicative group of unitary matrices called the {\em  Pauli group}. 
For a positive integer $n>1$, an {\em $n$-qubit Pauli matrix} can be represented as
\[X^uZ^v=X^{u_1}Z^{v_1}\otimes X^{u_2}Z^{v_2}\otimes \cdots \otimes X^{u_n}Z^{v_n} \in \mathcal{H}^{2^n}, \]
where $u=(u_1,u_2,\ldots,u_n),v=(v_1,v_2,\ldots,v_n)\in \F_2^n$. In particular, the matrix $X^uZ^{v}$ acts on the $i$-th qubit as $X^{u_i}Z^{v_i}$ for each $1\le i \le n$. 
For example, $X^{(1,0,1,0)}Z^{(0,0,1,1)}$ is equivalent to the Pauli matrix $X\otimes I \otimes XZ \otimes Z.$
Pauli matrices are also referred to as Pauli errors, as they are commonly considered as the foundational error model encompassing all potential errors capable of corrupting a quantum system. 
This is a consequence of the so-called error discretization that results from the syndrome extraction process, where arbitrary continuous errors discretize in such set of discrete errors \cite{KLdisc}.
The weight of an $n$-qubit Pauli matrix is the
number of non-identity components in its tensor product.
Analogously, the group generated by the $n$-dimensional Pauli matrices with the multiplicative factors $\{\pm 1,\pm i\}$, $P_n=\displaystyle\bigotimes^{n}P_1$, is called the {\em $n$-dimensional Pauli group}. 
% The normalizer group of $P_n$ in $U(n)$, the group of all $n\times n$ unitary matrices over $\mathcal{H}$, is called the {\em Clifford group} of $n$ qubits and we denote it by $\Cl_n$. One can show that $\Cl_n$ can be generate by the tensor product of the following gates: 
% \begin{center}
% $S=\sqrt{Z}=\begin{bmatrix}1&0\\0&i\end{bmatrix}$, 
% $H=\frac{1}{\sqrt{2}}\begin{bmatrix}1&1\\1&-1\end{bmatrix}$, 
% and $^CX=
% \begin{bmatrix}1&0&0&0\\
% 0&1&0&0\\
% 0&0&0&1\\
% 0&0&1&0
% \end{bmatrix}$.
% \end{center}
% However, the group $\Cl_n$ is not sufficient for universal quantum computation. In fact it can be shown that a quantum
% computer using only operations from $\Cl_n$ can be efficiently simulated on a
% classical computer. It is well-known
% that $\Cl_n$ augmented by any matrix in $U(n) \setminus \Cl_n$, can approximate any other unitary operator arbitrarily good. A common choice for such non-Clifford gate is $$T=\sqrt{S}=\begin{bmatrix}1&0\\0&e^{i\frac{\pi}{4}}\end{bmatrix}.$$

Recall that an $[\![n,k,d]\!]_2$ binary quantum code encodes $k$ qubits of information into $n$ physical qubits (a $2^k$-dimensional subspace of $\mathcal{H}^{2^n}$) and is capable of correcting up to $\lfloor \frac{d-1}{2} \rfloor$ or fewer number of errors on the $n$ physical qubits.

The following theorem describes the necessary conditions to construct a subclass of quantum codes known as {\em quantum stabilizer codes}, and also to determine their parameters. For any subgroup $G$ of $P_n$, we denote the Pauli normalizer of $G$ by $N(G)$, which is the set of Pauli matrices that commute with each element of $G$.

\begin{theorem}\label{actual quantum}
Let $G$ be a commutative subgroup of $P_n$ not containing a non-trivial multiple of the identity that is generated by a minimal set of $n-k$ Pauli matrices. Then, the common eigenspace with eigenvalue 
+1 of $G$ is a $[\![ n,k, d]\!]_2$ binary quantum code. If $k>0$, then
$d$ is the minimum weight of error operators in $N(G) \setminus G$; if $k=0$, then 
 $d$ is the minimum weight of error operators in $N(G)=G$.
\end{theorem}

\begin{proof}
For the proof, see, for example, the discussion in Section~2.2 of \cite{Panteleev2021degeneratequantum} or Section~3.2 of \cite{Grassllinking}.    
\end{proof}

A quantum code of Theorem \ref{actual quantum} is called {\em degenerate} if it minimum distance $d$ is greater than the minimum weight of error operators in $N(G)$. 
In general, directly constructing a quantum code or computing its parameters using the description provided in Theorem \ref{actual quantum} is not practical. 
Instead, a more feasible approach to constructing stabilizer codes is through the Calderbank-Shor-Steane (CSS) construction, as detailed below.

\begin{theorem}\cite[Theorem 9]{Calderbank}\label{T:CSS}
Let $C_2 \subseteq C_1$ be binary linear codes of length $n$ with dimensions $k_2$ and $k_1$, respectively. Then there exists an $[\![n,k_1-k_2,d]\!]_2$ binary quantum stabilizer code, where 
\begin{equation}\label{E: degeneracy}
d=\min\{d(C_1 \setminus C_2), d(C_2^\bot \setminus C_1^\bot)\}.
\end{equation}
\end{theorem}

In particular, the stabilizer group of the quantum code described in Theorem \ref{T:CSS} is generated by 
\[\{X^u:u \in C_2\} \cup \{Z^v:v \in C_1^\bot\}.\] 
Moreover, this quantum code is degenerate if $d>\min\{d(C_1), d(C_2^\bot)\}.$
The quantum codes discussed in this paper are generated by applying the CSS construction to binary duadic and triadic codes. 
Our primary interest is in computing the parameters and identifying the degenerate codes constructed in this manner.
Since Theorem \ref{T:CSS} completely determines the parameters and degeneracy of quantum CSS codes, we never deal with the actual quantum code in the sense of Theorem \ref{actual quantum}.

A special case of CSS construction arises when the pair of binary codes satisfy $C_1=C_2^\bot$ and both codes are cyclic. As discussed in \cite{grasslcyclic,Steane_1999}, such quantum CSS codes support transversal Pauli operators, the $S=Z^{\frac{1}{2}}$ gate, and the Hadamard $H$ gate (bitwise application of such operators maps the encoded Hilbert space
onto itself). 
Indeed, many of the quantum CSS codes in this paper satisfy the mentioned property.

%%%%%%%%%%%%%%%%%%%%%%%%%%%%%%%%%%%%%%%%%%%%%
\section{Existence of Duadic Codes and Quantum CSS Duadic Codes} \label{S:Existence}
%%%%%%%%%%%%%%%%%%%%%%%%%%%%%%%%%%%%%%%%%%%%%

In this section, we investigate the existence of binary duadic codes (Definition \ref{D:duadic}). 
Duadic codes with splittings satisfying $|X|=1$ have already been discussed in the literature; see, for example, \cite{aly2006,RezaDuadic}. 
Our focus and new results lie in exploring the existence of splittings satisfying $|X|>1$ and their corresponding binary CSS codes. This latter case has not been previously explored in the literature.

First note that the Euclidean dual-containing duadic codes are minimal among all Euclidean dual-containing cyclic codes in the sense of the following proposition. In particular, the odd-like (respectively even-like) duadic codes have the largest minimum distance among all dual-containing (respectively self-orthogonal) cyclic codes. 
\begin{proposition}
Let $C$ be a binary dual-containing linear cyclic code of length $n$. Then 
\begin{equation}\label{E:Optimally}
C^\bot \subseteq D_1 \subseteq C_1=D_1^\bot \subseteq C,
\end{equation}
where $D_1$ and $C_1$ are even-like and odd-like duadic codes corresponding to a splitting that is given by $\mu_{-1}$, respectively.
\end{proposition}

\begin{proof}
The proof is similar to that of \cite[Proposition 3.1]{DuadicConsta}, thus, we only give a sketch of it.  Let $A$ be the defining set of $C$. By Theorem \ref{T:Euclidean}, we have $A \cap \mu_{-1}(A)=\emptyset$. Hence we can find a splitting of $\Z/n\Z$ in the form $(X,S_1,S_2)$ that is given by $\mu_{-1}$, where $A\subseteq S_1$, respectively. Let $D_1$ and $C_1$ be even-like and odd-like duadic codes with the defining sets $S_1 \cup X$ and $S_1$. Then $D_1 \subseteq C_1=D_1^\bot$ and $C_1 \subseteq C$. 
Now the fact that $C$ is dual-containing implies (\ref{E:Optimally}).
\hfill $\square$ \end{proof}
According to the proposition above, among all quantum CSS codes constructed using a dual-containing cyclic code and its dual, the duadic pairs yield the largest minimum distance.
Additionally, the existence of duadic codes relies on the existence of a splitting for $\mathbb{Z}/ n\mathbb{Z}$. 
Furthermore, $\mu_{b}(\{0\})=\{0\}$ for each multiplier $\mu_b$.
So in the next lemmas, we study when $a \in \Z/n\Z$ satisfies $Z(a)=Z(ba)$. The proofs are straightforward, but we state them for the sake of completeness.
Recall that $\ord_n(2)=|Z(1)|$ is the multiplicative order of $2$ modulo $n$. 

\begin{lemma}\label{L:1}
Let $n$ and $b$ be positive odd integers such that $\gcd(n,b)=1$. Then $Z(s)=Z(bs)$ for each $s \in \Z/n\Z$ if and only if $n\mid 2^r-b$ for some $1\le r \le \ord_n(2)$.    
\end{lemma}

\begin{proof}
$\rightarrow:$ The assumption implies that $Z(1)=Z(b)$ (choosing $s=1$). Thus there exists $1\le r \le \ord_n(2)$ such that $2^r \equiv b \pmod{n}$, or equivalently $n \mid 2^r-b$.\\
$\leftarrow:$ We have $2^r \equiv b \pmod n$. Now multiplying both sides by $s$ gives the desired result. 
\hfill $\square$ 
\end{proof}

Therefore, if $n\mid 2^r-b$ for some integer $1\le r \le \ord_n(2)$, then there is no splitting of $\Z /n \Z$ that is given by $\mu_b$. 
On the other hand, if $n\nmid 2^r-b$ for each integer $1\le r \le \ord_n(2)$, there may still exist some values  $s \in \Z/n\Z$ for which the equality $Z(s)=Z(bs)$ holds. 
The trivial case occurs when $s=0$.
The following lemma classifies all other fixed cyclotomic cosets of $\Z/n\Z$ under the action of a multiplier $\mu_{b}$. 

\begin{lemma}\label{L:2}
 Let $n$ and $b$ be positive odd integers such that $\gcd(n,b)=1$, and $s\in \Z/n\Z$. Then $Z(s)=Z(bs)$ if and only if $\frac{n}{\gcd(n,2^r-b)}\mid s$ for some $1\le r \le \ord_n(2)$.   
\end{lemma}

\begin{proof}
We have
$Z(s)=Z(bs)$ if and only if $2^rs\equiv bs \pmod{n}$ or equivalently $n\mid s(2^r-b)$ for some $1\leq r \leq \ord_n(2)$. 
The latter is equivalent to $\frac{n}{\gcd(n,2^r-b)}\mid s$.
\hfill $\square$ \end{proof}

For positive integers $n$ and $b$ such that $\gcd(n,b)=1$, we define 
\begin{equation}\label{E:fix}
\Fix_b(n) =\{ s \in \Z/ n\Z: \frac{n}{\gcd(n,2^r-b)}\mid s \ \text{for some} \ 1\le r \le \ord_n(2) \}.
\end{equation}
Note that, by Lemma \ref{L:2}, the set $\Fix_b(n)$ is a union of $2$-cyclotomic cosets modulo $n$. Also, $Z(s)=Z(bs)$ if and only if $s \in \Fix_b(n)$. Moreover, Lemma \ref{L:1} implies that $\Fix_b(n)=\Z/n\Z$ if and only if $n\mid 2^r-b$ for some $1\le r \le \ord_n(2)$. This leads to the following observation. 

\begin{theorem}\label{T:Duadic existence}
 Let $n$ be a positive odd integer. \\
 (1) Dual-containing binary duadic codes of length $n$ exist if and only if 
 $\Fix_{-1}(n) \subsetneq \Z/n\Z$. \\
 (2) Each splitting of a duadic code that is given by $\mu_{b}$ is in the form $(X,S_1,S_2)$, where $X=\Fix_{b}(n)$. \\
 (3) $\Fix_{b}(n)=\mu_{-1}(\Fix_{b}(n))$. 
\end{theorem}

\begin{proof}
First, using Lemma \ref{L:1}, one can argue that dual-containing binary duadic codes exist if and only if there exists a splitting of $\Z/n\Z$ that is given by $\mu_{-1}$ if and only if $\Fix_{-1}(n) \subsetneq \Z/n\Z$. 

For the second part, note that $\Fix_{b}(n)\subseteq X$. Moreover, the definition of a splitting, Definition \ref{splitting def}, implies that $X \subseteq \Fix_{b}(n)$. Thus, $X=\Fix_{b}(n)$.

The last part follows from (\ref{E:fix}) as for each $s\in \Z/n\Z$, we have $\frac{n}{\gcd(n,2^r-b)}\mid s$ if and only if $\frac{n}{\gcd(n,2^r-b)}\mid -s$.
\hfill $\square$ \end{proof}

Now we use the above information to formally introduce quantum CSS duadic codes. 

\begin{theorem}\label{T:quantum duadic}
Let $n$ and $b$ be positive odd integers such that $\gcd(n,b)=1$ and $(\Fix_b(n),S_1,S_2)$ is a splitting of $\Z/n\Z$ that is given by $\mu_b$. Then there exists a binary quantum code with parameters $[\![n,|\Fix_b(n)|,d\ge \sqrt{d(C)}]\!]_2$, where $d(C)$ is the minimum distance of binary cyclic code with defining set $(\Z/n\Z) \setminus \Fix_b(n)$.
\end{theorem}

\begin{proof}
Let $D_1$ and $C_1$ be binary even-like and odd-like duadic codes of length $n$ with the defining sets $\Fix_b(n) \cup S_1$ and $S_1$, respectively. Then by Theorem \ref{T:Duadic-properties} (2), we have   $\dim(C_1)=\frac{n+|\Fix_b(n)|}{2}$ and $\dim(D_1)=\frac{n-|\Fix_b(n)|}{2}$. 
Moreover, Theorem \ref{T:Duadic existence} implies that $\mu_{-1}(\Fix_b(n))=\Fix_b(n)$. So we have that $(\Fix_b(n),-S_1,-S_2)$ is also a splitting of $\Z/n\Z$ that is given by $\mu_{b}$. 
Furthermore, $C_1^\bot$ and $D_1^\bot$ are  binary even-like and odd-like duadic codes of length $n$ with the defining sets $\Fix_b(n) \cup -S_2$ and $-S_2$, respectively (the codes $D_1$ and $C_1^\bot$, similarly $C_1$ and $D_1^\bot$, are permutation equivalent as $\mu_{-b}$ exchanges their defining sets).
Now, Theorem \ref{T:Duadic-properties} (5) implies that 
\[d(C_1 \setminus D_1) \ \text{and}\ d(D_1^\bot \setminus C_1^\bot)\ge \sqrt{d(C)}. \] 
Applying the CSS construction (Theorem \ref{T:CSS}) to $D_1 \subseteq C_1$, proves the existence of an $[\![n,\Fix_b(n),d \ge 
\sqrt{d(C)}]\!]_2$. 
\hfill $\square$ \end{proof}

The mentioned theorem presents a generalization of the conventional definition of binary quantum CSS duadic codes as outlined in \cite{aly2006}. 
This is because when $X=\{0\}$, the code $C$ in Theorem \ref{T:quantum duadic} has the minimum distance $n$ ($C$ is the code generated by all-one). This scenario gives the family of quantum CSS duadic codes that has parameters $[\![n,1,d \ge \sqrt{n}]\!]_2$ as established in \cite[Theorem 4]{aly2006}.

Note also that in many cases minimum distance of the code $C$ in Theorem \ref{T:quantum duadic} can be computed or bounded using the BCH bound.
In particular, if $s$ is the smallest non-zero element of $\Fix_b(n)$, then one can easily show that $[1,s-1] \subseteq (\Z/n\Z)\setminus \Fix_b(n)$. Therefore, applying the BCH lower bound implies that $d(C) \ge s$.
For example, when $n=21$, the code $C$ has the defining set $(\Z/21\Z) \setminus \{0,7,14\}$ which contains the interval $[1,6]$. Hence it has minimum distance $d(C) \ge 7$. In fact $7$ is the minimum distance in this case.

%%%%%%%%%%%%%%%%%%%%%%%%%%%%%%%%%%%%%%%%%%%%%%%%%%%%%%%%%%%%%%%%%%
\section{Extended Splitting and Degenerate Codes}\label{S:Extended splitting}

In this section, we study binary duadic codes that are constructed by extending the splittings of shorter-length binary duadic codes. This approach allows us to construct degenerate quantum duadic codes.
We also provide lower and upper bounds, or even compute, the minimum odd-like distance of duadic codes constructed in this manner. To begin, we first define the extended splitting of duadic codes.

\begin{definition}\label{D:extended Splitting}
Let $n_1$ and  $n_2$ be positive odd integers, and $T=(T_0,T_1,T_2)$ and $U=(U_0,U_1,U_2)$ be two splittings of $\Z/n_1\Z$ and $\Z/n_2\Z$, respectively, that are given by $\mu_b$ such that $U_0=\{0\}$. The set $S=(S_0,S_1,S_2)$, 
where 
$S_0=\{i n_2: i \in T_0\}$
and 
\begin{equation}\label{E:S def}
    S_k=\{in_2:i \in T_k\} \cup \{i+jn_2: i \in U_k, 0\le j\le n_1-1\}
\end{equation}    
for $k=1,2$, is called the {\em extended splitting} of $T$ and $U$ modulo $n_1n_2$.
\end{definition}

When $T_0=\{0\}$, it has been shown in \cite{Smid} that the extended splitting of $T$ and $U$ is a splitting of $\Z/(n_1n_2\Z)$ that is given by $\mu_b$. 
The same conclusion is obtained in \cite{DuadicConsta} for cyclic and constacyclic duadic codes over $\F_4$ that are given by $\mu_{-2}$. 
A similar technique can be used to show that when $|T_0|>1$, the set $S$ is a splitting of $\Z/(n_1n_2\Z)$ that is given by $\mu_b$. 

Note also that each splitting of $\Z/ (n_1n_2)\Z$ is not necessarily an extended splitting of two smaller length splittings. 
For instance, there is no splitting of $\Z/3\Z$ and $\Z/5\Z$; however, one can find a splitting for $\Z/15\Z$.

The next theorem connects codewords of even-like duadic codes of smaller length and its corresponding extended duadic code. Proof of the following theorem is similar to that of Theorem 3 in \cite{Smid} and Theorem 4.8 in \cite{DuadicConsta}, and here we only provide a short proof. 
\begin{theorem}\label{T:Even distance}
 Let $C_S$ and $C_U$ be even-like binary duadic cyclic codes of lengths $n_1n_2$ and $n_2$ with defining sets $ S_0\cup S_1$ and $ U_0\cup U_1$, respectively, where $(S_0,S_1,S_2)$ is an extended splitting of a length $n_1$ splitting and a length $n_2$ splitting in the form $(U_0,U_1,U_2)$.
 If $C_U$ has a vector of weight $t$, then $C_S$ also has a vector of weight~$t$.
 In particular, 
 \[d(C_S)\le d(C_U)< n_2.\]
\end{theorem}

\begin{proof}
Let $\delta$ be a primitive $(n_1n_2)$-th root of unity such that the mentioned splittings are formed using the primitive roots $\delta$, $\delta^{n_2}$, and $\delta^{n_1}$.
First note that 
\[ n_1S_i=\{(n_1a) \mod n_1n_2: a \in S_i\}=
\begin{cases}
n_1U_i & \ \text{if} \ i= 1,2\\
\{0\} & \ \text{if} \ i=0.\\
\end{cases} \]
Hence, if $c(x) \in C_U$ (or equivalently $c(\delta^{n_1a})=0$ for each $a \in U_0 \cup U_1$), then $c'(x)=c(x^{n_1}) \in C_S$. This is because 
for any $b \in S_0\cup S_1$, we have $c'(\delta^b)=c(\delta^{n_1b})$ and as we showed above $n_1b\in (n_1S_0)\cup (n_1S_1)=n_1U_0 \cup n_1U_1$ (note that $n_1S_0=n_1U_0=\{0\}$). Therefore, $c'(\delta^ b)=0$. Thus, 
$c'(x) \in C_S$ and it has the same weight as $c(x)$. The last part is an immediate consequence of the first one. 
\hfill $\square$ \end{proof}

Let $T$ and $U$ be two splittings of length $n_1$ and $n_2$ that are given by $\mu_{a}$ and $\mu_b$, where $\gcd(n_1,n_2)=1$. Then, using the Chinese remainder theorem, one can find $c \in \Z/(n_1n_2)\Z$ such that $c \equiv a \pmod{n_1}$ and $c \equiv b \pmod{n_2}$. Thus the set $S$ defined in Definition \ref{D:extended Splitting} is still an splitting that is given by $\mu_c$. 
Therefore, the result of Theorem \ref{T:Even distance} holds in such cases too.

An application of the above theorem is in the construction of degenerate quantum CSS duadic codes with dimension greater than one. 

\begin{corollary}\label{C:deg duadic}
Let $n_1$ and $n_2$ be positive odd integers, and let $T=(T_0,T_1,T_2)$ and $U=(U_0,U_1,U_2)$ be two splittings of length $n_1$ and $n_2$, respectively, that are either given by the same multiplier, or they have different multipliers and $\gcd(n_1,n_2)=1$. If $\frac{n_1}{|T_0|}\ge n_2$, then there exists a degenerate binary quantum code with parameters $[\![n_1n_2,|T_0|,d> n_2]\!]_2$.
\end{corollary}

\begin{proof}
Let $S=(S_0,S_1,S_2)$ be the extended splitting of $T$ and $U$ that is given by $\mu_b$ for some $b \in \Z /(n_1n_2)\Z$. 
Let $C_S$ and $C_U$ (respectively $D_S$ and $D_U$) be even-like (respectively odd-like) binary duadic codes of lengths $n_1n_2$ and $n_2$ with the defining sets $S_0 \cup S_1$ and $U_0 \cup U_1$ (respectively $S_1$ and $U_1$), respectively.
Then by Theorem \ref{T:Even distance} we have $d(C_S)\le d(C_U) \le n_2$.  

Moreover, the fact that $\frac{n_1}{|T_0|}\ge n_2$ implies that there exist integers $a<b \in T_0$ such that $b-a\ge n_2-1$ and $T_0 \cap \{a,a+1,\ldots,b\}=\emptyset$. Hence the the set $S_0$ has no element in the interval $[n_2a,n_2b]$ which has size $n_2^2$. Let $C$ be the cyclic code of length $n_1n_2$ and the defining set $A=\big(\Z/(n_1n_2)\Z \big)\setminus S_0$. 
Then $[n_2a,n_2b] \subseteq A$, and the BCH bound (Theorem \ref{T:BCH}) implies that $d(C) > n_2^2$. 
Now, applying the result of Theorem \ref{T:quantum duadic} to the duadic codes corresponding to $S$ (recall that $\Fix_b(n_1n_2)=S_0$ and $|S_0|=|T_0|$) implies the existence of a binary quantum code 
$Q$ with parameters $[\![n_1n_2,|T_0|,d> n_2]\!]_2$, where $Q$ is the CSS code of $C_S \subset D_S$. Moreover, Theorem \ref{T:Even distance} implies that $d(C_S)< n_2$ and since $C_S$ and $D_S^\bot$ are permutation equivalent (see the proof of Theorem \ref{T:quantum duadic}) we have $d(D_S^\bot)< n_2$. Therefore, $Q$ is degenerate. 
\hfill $\square$ \end{proof}

A special case of above theorem is when $T_0=\{0\}$ which is discussed in \cite[Theorem 6]{aly2006} in more detail. 
A remaining problem is to compute or bound the  minimum distance of such quantum codes using the odd-like distance of smaller duadic codes. In what follows we resolve this problem. 

The next theorem gives a bound for the minimum odd-like distance of duadic codes that have been constructed using the extended splittings. 
\begin{theorem}\label{T:duadic distance}
 Let $n_1$ and $n_2$ be positive odd integers.
 Let $C$, $C_1$, and $C_2$ be odd-like binary duadic codes of lengths $n_1n_2$, $n_1$, and $n_2$, respectively, where $C$ is constructed from an extended splitting of $C_1$ and $C_2$. 
Then 
\[d_o(C_1)d(C_2) \le d_o(C)  \le d_o(C_1)d_o(C_2) .\]
In particular, if $d(C_2)=d_o(C_2)$, we have $d_o(C)=d_o(C_1)d_o(C_2).$
\end{theorem}

\begin{proof}
 The proof follows the logic of \cite[Theorem 4.11]{DuadicConsta}, and we omit it here. 
\hfill $\square$ \end{proof}

This result provides an efficient way of computing or bounding the minimum odd-like distance of binary duadic codes of certain composite lengths. Note also that when the code $C_2$ in the above theorem is a QR code, then $d(C_2)=d_o(C_2)$ is odd \cite[Theorem 6.6.22]{Huffman}, so this bound gives the exact minimum odd-like distance of such codes.  

\begin{corollary}
 Let $C$, $C_1$, and $C_2$ be as in Theorem \ref{T:duadic distance} and $C_2$ be a QR code. Then  $d_o(C)=d_o(C_1)d_o(C_2).$   
\end{corollary}

For instance the binary odd-like QR code of length $7$ which has parameters $[7,4,3]_2$ can be used to construct a binary quantum code with parameters $[\![7,1,3]\!]_2$. This code has minimum even-like distance of $4$. Now repeatedly extending the splittings and using the above distance bound, one get the family of $[\![7^m,1,3^m]\!]_2$ quantum codes for each $m\ge 1$. 
Except for the case $m=1$, the remaining codes are all degenerate, and their corresponding duadic codes always have minimum distance 4. 
This observation completes the discussion in Example 7 of \cite{aly2006} by computing the exact minimum distance of such codes.

We conclude this section by presenting another example and a list of quantum CSS duadic codes. 
The parameters of all small-length quantum codes discussed in this paper were computed using the computer algebra system Magma \cite{magma}.

\begin{example}
One can find a splitting of $\Z/15\Z$ that is given by $\mu_{-1}$ in the form $T=(T_0=Z(0) \cup Z(3) \cup Z(5),T_1=Z(1),T_2=Z(7))$. The quantum duadic code of length $15$ corresponding to this splitting has parameters $[\![15,7,3]\!]_2$. Moreover, there exists a splitting of $\Z / 7\Z$ in the form $U=(U_0=Z(0),U_1=Z(1),U_2=Z(3))$ that is given by $\mu_{-1}$. The splitting $U$ gives the  $[\![7,1,3]\!]_2$ quantum code discussed above. 
Extending the splittings $T$ and $U$, one gets a splitting of $\Z/105\Z$ in the form 
\[
\begin{split}
S=(S_0=Z(0)\cup Z(21)& \cup Z(35), S_1=Z(1)\cup Z(7)\cup Z(9)\cup Z(11)\cup Z(15)\cup Z(25),\\&S_2=Z(3)\cup Z(5)\cup Z(13)\cup Z(17)\cup Z(45)\cup Z(49)).
\end{split}
\]
Now, Corollary \ref{C:deg duadic} and Theorem \ref{T:duadic distance} imply that the quantum duadic corresponding to the splitting $S$ has parameters $[\![105,7,9]\!]_2$ which is degenerate, and pure to $4$.
Further extending the splittings $S$ and $U$, one gets a $[\![7^2 \times 3 \times 5, 7, 3^3]\!]_2$ binary quantum code that is degenerate and pure to 4. Repeating this process one gets an infinite family of quantum codes 
with parameters 
$[\![7^i \times 3 \times 5, 7, 3^{i+1}]\!]_2$
which is again degenerate and pure to 4. 
\end{example}
Hence, one can generate many families of degenerate quantum codes with dimensions larger than one and arbitrarily large minimum distances using binary duadic codes. 

Table \ref{TL:1} presents the parameters of some additional small-length quantum duadic codes with dimension greater than one. In the table, the coset leaders are those of the odd-like duadic codes and 
all the splitting are given by $\mu_{-1}$. According to \cite[Theorem 2.1]{grasslcyclic}, the weights of stabilizer elements (even-like duadic codes) are all divisible by $4$. 
Therefore, as mentioned earlier, these codes support transversal Pauli operators, the $S=Z^{\frac{1}{2}}$ gate, and the Hadamard $H$ gate.

It should be also mentioned that some of these codes offer more logical qubits compared to the best quantum (non-CSS) duadic codes constructed using duadic codes over $\F_4$ \cite[Table 1]{DuadicConsta}. 
For example, the quaternary duadic of length $45$ yields a quantum code with parameters $[\![45,9,5]\!]_2$, whereas the binary duadic code of the same length results in a quantum code with parameters $[\![45,13,5]\!]_2$, providing four additional logical qubits.

Furthermore, each $[\![n,k,d]\!]_2$ quantum code listed in Table~\ref{TL:1} with an odd minimum distance $d$ can be used to generate a quantum code with parameters $[\![n+1,k-1,d+1]\!]_2$. 
This is achieved by employing the CSS construction on the extended code of odd-like duadic code and its dual. 

\begin{table}[ht]
\captionsetup{justification=centering}
\caption{Small-length quantum duadic code of dimension larger than one.} 
\label{TL:1}
\begin{center}
\scalebox{1}{
\begin{tabular}{ |p{1.3 cm}|p{3.1 cm}| p{2.0 cm}| p{2.0 cm}|}
\hline
 Length & Coset Leaders &Parameters& Degenerate \\
 \hline
$15$ & $1$                 &  $[\![15,7,3]\!]_2$  & No \\
$21$ & $1,3$               &  $[\![21,3,5]\!]_2$  & No \\
$35$ & $1,5$               & $[\![35,5,6 ]\!]_2$ & No \\
$45$ & $1,3$               & $[\![45,13,5]\!]_2$& No \\
$55$ & $1$                 & $[\![55,15,5]\!]_2$& No \\
$85$ & $1,3,7,9$           & $[\![85,21,5]\!]_2$& No \\
$91$ & $1,3,9,13$          & $[\![91,13,7]\!]_2$& No \\
$93$ & $1,5,7,21,33,45$      & $[\![93,3,14]\!]_2$& No\\
$95$ & $1$                 &$[\![95,23,5]\!]_2$ & No \\
$105$ & $3,5,7,11,13,15$  & $[\![105,7,12]\!]_2$ & Yes \\
$115$ & $1,5$  & $[\![115,5,14]\!]_2$ & No \\
\hline
\end{tabular}}
\end{center}
\end{table}

%%%%%%%%%%%%%%%%%%%%%%%%%%%%%%%%%%%%%%%%%%%%%%%%%%
\section{Quantum Triadic Codes}\label{S:Triadic}
%%%%%%%%%%%%%%%%%%%%%%%%%%%%%%%%%%%%%%%%%%%%%%%%%%%

Triadic codes, introduced in \cite{pless1988triadic}, represent another infinite class of cyclic codes with many intriguing properties \cite{Pless3,Triadic3}. 
In this section, we provide a brief overview of these codes before proceeding to construct a family of binary quantum codes using them.

Let $n$ be a positive odd integer. The tuple $(X_\infty,X_0,X_1,X_2)$ is called a {\em 3-splitting} of $\Z/n\Z$ if 
\begin{itemize}
    \item $X_i$ is a union of 2-cyclotomic cosets modulo $n$ for each $i=0,1,2,\infty$.
    \item $(X_\infty,X_0,X_1,X_2)$ is a partition of  $\Z/n\Z$.
    \item There exists a multiplier $\mu_{b}$ such that $\mu_{b}(X_0)=X_{1}$, $\mu_{b}(X_1)=X_{2}$, $\mu_{b}(X_2)=X_{0}$, and $\mu_{b}(Z(s))=Z(s)$ for each $s \in X_\infty$.
\end{itemize}
Note that despite the duadic case, $\mu_b$ cannot be an involution, i.e., $b^2 \not\equiv 1 \pmod n $. 
\begin{definition}
Let $(X_\infty,X_0,X_1,X_2)$ be a 3-splitting of $\Z/n\Z$ that is given by $\mu_{b}$. The binary linear cyclic codes $C_1$ and $C_2$ of length $n$ with the defining sets $X_i$ and $X_\infty\cup X_i\cup X_{j}$, where $i\ne j \in\{0,1,2 \}$, are called a pair of odd-like and even-like binary {\em triadic} codes of length $n$. 
\end{definition}

Some properties of triadic codes is summarized below.

\begin{proposition}\label{P:Triadic properties}
 Let $C_1$ and $C_2$ ($C_2\subset C_1$) be a pair of odd-like and even-like binary triadic codes corresponding to a 3-splitting that is given by multiplier $\mu_{b}$. Then
 \begin{enumerate}
     \item $C_1$ and $\mu_b(C_1)$ (similarly $C_2$ and $\mu_{b}(C_2)$) are a pair of permutation equivalent triadic codes. 
     \item $\dim(C_1)=\frac{2n+|X_\infty|}{3}$ and $\dim(C_2)=\frac{n-|X_\infty|}{3}$.
     \item The codes $D_1=C_2^\bot$ and $D_2=C_1^\bot$ form a pair of odd-like and even-like triadic codes, and $d(C_1 \setminus C_2)=d(D_1 \setminus D_2)$.
%      \item If $\mu_{-1}(X_t)=X_t$ for each $t \in \{0,1,2,\infty\}$, then the codes $D_1=C_2^\bot$ and $D_2=C_1^\bot$ form another pair of odd-like and even-like triadic codes, and $d(C_1 \setminus C_2)=d(D_1 \setminus D_2)$.
 \end{enumerate}
 \end{proposition}
\begin{proof}
 We only discuss the last part, as the proof for the other cases is straightforward. Let $(X_\infty,X_0,X_1,X_2)$ be a 3-splitting of $\Z/n\Z$ that is given by $\mu_{b}$, and $X_i$ and $X_\infty\cup X_i\cup X_{j}$ be the defining sets of $C_1$ and $C_2$ for some $ i\ne j \in\{0,1,2 \}$, respectively. 
 A similar proof as in Theorem \ref{T:Duadic existence} shows that $\mu_{-1}(X_\infty)=X_\infty$. Thus $D_2=C_1^\bot$ has the defining set
 \[(\Z/n\Z)\setminus (-X_i)=X_\infty\cup -X_{j}\cup -X_{k},\]
 where $k \in \{0,1,2\} \setminus\{i,j\}$.
 Similarly $D_1=C_2^\bot$ has the defining set
 \[(\Z/n\Z)\setminus (X_\infty\cup -X_i\cup -X_{j})= -X_{k}.\]
 Therefore $D_1$ and $D_2$ are another pair of odd-like and even-like triadic codes (corresponding to the splitting $(X_\infty,-X_0,-X_1,-X_2)$). 
 
For the second argument, note that we can find multiplier $b' \in \{b,b^2\}$ such  that $\mu_{-b'}(X_i)=-X_k$ and $\mu_{-b'^2}(X_\infty\cup X_i\cup X_{j})=X_\infty\cup -X_j\cup -X_{k}$. Thus the codes $C_1$ and $D_1$, respectively $C_2$ and $D_2$, are permutation equivalent and therefore have the same weight distribution. Now the facts that $C_2 \subset C_1$ and $D_2 \subset D_1$ imply that the sets $C_1 \setminus C_2$ and $D_1 \setminus D_2$ have the same weight distributions. This proves the last statement.
% For the second argument, first note that we can find a multiplier $b' \in \{b,b^2\}$ such  that $\mu_{b'}(C_1)=D_1$ and $\mu_{b'^2}(C_2)=D_2$.
% Let $c$ be a minimum weight vector in $C_1 \setminus C_2$. If $\mu_{b'}(c) \in D_1 \setminus D_2$ then $d(C_1 \setminus C_2) \ge d(D_1 \setminus D_2)$. So assume that $\mu_{b'}(c)\in D_2$. Since $\mu_{b'}(D_2)=C_2$, we have $\mu_{b'^2}(c) \in C_2 \subseteq C_1$. This implies that $c=\mu_{b'^3}(c) \in D_1$. However, $c \notin D_2$ as otherwise $c \in D_2 \cap C_1=\{0\}$. Thus again $d(C_1 \setminus C_2) \ge d(D_1 \setminus D_2)$. 
%A similar approach can be applied to show that $d(C_1 \setminus C_2) \le d(D_1 \setminus D_2)$. 
%Therefore, $d(C_1 \setminus C_2)=d(D_1 \setminus D_2)$.
 \hfill $\square$ \end{proof}

For certain values of $n$ (but not always), duadic codes of length $n$ may not exist; however, triadic codes can be constructed. The following example illustrates this.

\begin{example}
Let $n=43$. Then $\Z/n\Z=Z(0) \cup Z(1) \cup Z(3) \cup Z(7)$, where $|Z(1)|=|Z(3)|=|Z(7)|$. Moreover for any positive integer $1\le b \le 42$, the multiplier $\mu_b$ either fixes all $Z(1),Z(3),Z(7)$ or fixes none of them. Therefore, there is no $2$-splitting modulo $43$. However, $(Z(0),Z(1),Z(3),Z(7))$ forms a 3-splitting that is given, for example, by $\mu_{3}$. 
\end{example}

In the next theorem, we discuss the structure of binary quantum triadic codes.  

\begin{theorem}\label{T:Triadic}
Let $n$ be a positive odd integer, and  $(X_\infty,X_0,X_1,X_2)$ be a 3-splitting of $\Z/n\Z$. Then there exists a binary quantum code with parameters $[\![n,\frac{n+2|X_\infty|}{3},d]\!]_2$, 
where 
$d=d(C_1\setminus C_2)$
and $C_1$ and $C_2$ are a pair of odd-like and even-like triadic codes, respectively.
\end{theorem}

\begin{proof}
Let $C_1$ and $C_2$ be a pair of odd-like and even-like triadic codes, respectively. Then $C_2 \subset C_1$ and by Proposition \ref{P:Triadic properties} (2) we have $\dim(C_1)-\dim(C_2)=\frac{n+2|X_\infty|}{3}$. Moreover, Proposition \ref{P:Triadic properties} (3) implies that $d(C_1 \setminus C_2)=d(C_2^\bot \setminus C_1^\bot)$. Hence applying the quantum CSS construction to $C_2 \subset C_1$ implies the desired result. 
\hfill $\square$ \end{proof}

Note that if $p$ is a prime number such that $3 \mid p-1$ and $2$ is a cubic residue modulo $p$, then there exists a 3-splitting of $\Z/p\Z$ \cite{Job}. 
In this case, we have $X_\infty=\{0\}$, and the corresponding quantum triadic code has rate $\frac{n+2}{3n}$. 
The quantum triadic family of Theorem \ref{T:Triadic} has a non-zero (asymptotic) rate of $\ge \frac{1}{3}$. Let $C_1$ and $C_2$ be a pair of odd-like and even-like triadic codes of length $n$.  Note also that when $C_1$ has an odd minimum distance, then $d(C_1 \setminus C_2)=d(C_1)$ (because $C_2$ only has even weight codewords). Thus one can find the cubic root minimum distance lower bound $d(C_1) \ge n^{\frac{1}{3}}$ using the discussion of \cite[Theorem 16]{Pless3}. 
Such odd minimum distance triadic codes form a family of binary quantum triadic codes with a non-zero asymptotic rate and growing minimum distance.

Next we look at the smallest example of a quantum triadic code. 
\begin{example}
Let $n=31$. Then 
$$(X_\infty=Z(0),X_0=Z(1)\cup Z(15), X_1=Z(3)\cup Z(7),X_2=Z(5)\cup Z(15))$$ 
forms a splitting of $\Z/31\Z$ that is given by $\mu_{5}$.  Let $C_1$ and $C_2$ be a pair of binary odd-like and even-like triadic codes of length $n$ with the defining sets $X_0$ and $X_\infty \cup X_0 \cup X_1$ respectively. The codes $C_1$ and $C_2$ have parameters $[31,21,5]_2$ and $[31,10,10]_2$. Hence applying the result of Theorem \ref{T:Triadic} gives a binary quantum code with parameters $[\![31,11,5]\!]_2$. This code is also a quantum BCH code and it is considered as a good candidate for quantum error correction \cite{Steane_1999}.
\end{example}

\begin{table}[ht]
\captionsetup{justification=centering}
\caption{Small-length quantum triadic codes.} 
\label{TL:2}
\begin{center}
\scalebox{1}{
\begin{tabular}{ |p{1.3 cm}|p{3.0 cm}|p{1.6 cm}| p{2.2 cm}|}
\hline
 Length & Coset Leaders & Multiplier &Parameters \\
 \hline
$31$ & $1,3$ &$\mu_5$ &$[\![31,11,5]\!]_2$\\
$43$ & $1$ &$\mu_3$  &$[\![43,15,6]\!]_2$ \\
$93$ & $1,3,9,23$ &$\mu_5$  &$[\![93,33,7]\!]_2$\\
$109$ & $1$ &$\mu_3$  &$[\![109,37,10]\!]_2$\\
$127$ & $1,5,19,27,47,63$ &$\mu_3$ &$[\![127,43,13]\!]_2$ \\
$129$ & $1,3,19$ &$\mu_5$ &$[\![129,45,12]\!]_2$\\
$155$ & $1,5,23,75$ &$\mu_3$  &$[\![155,55,10]\!]_2$\\
\hline
\end{tabular}}
\end{center}
\end{table}

Table~\ref{TL:2} presents the parameters of quantum triadic codes of small lengths. In the table, the coset leaders are those of the odd-like triadic codes.

\begin{figure}[ht]
    \caption{Minimum distance of quantum triadic codes as a function of the code length.}
    \label{f:quantum_codes_graph}
    \centering
\includegraphics[width=0.5\textwidth]{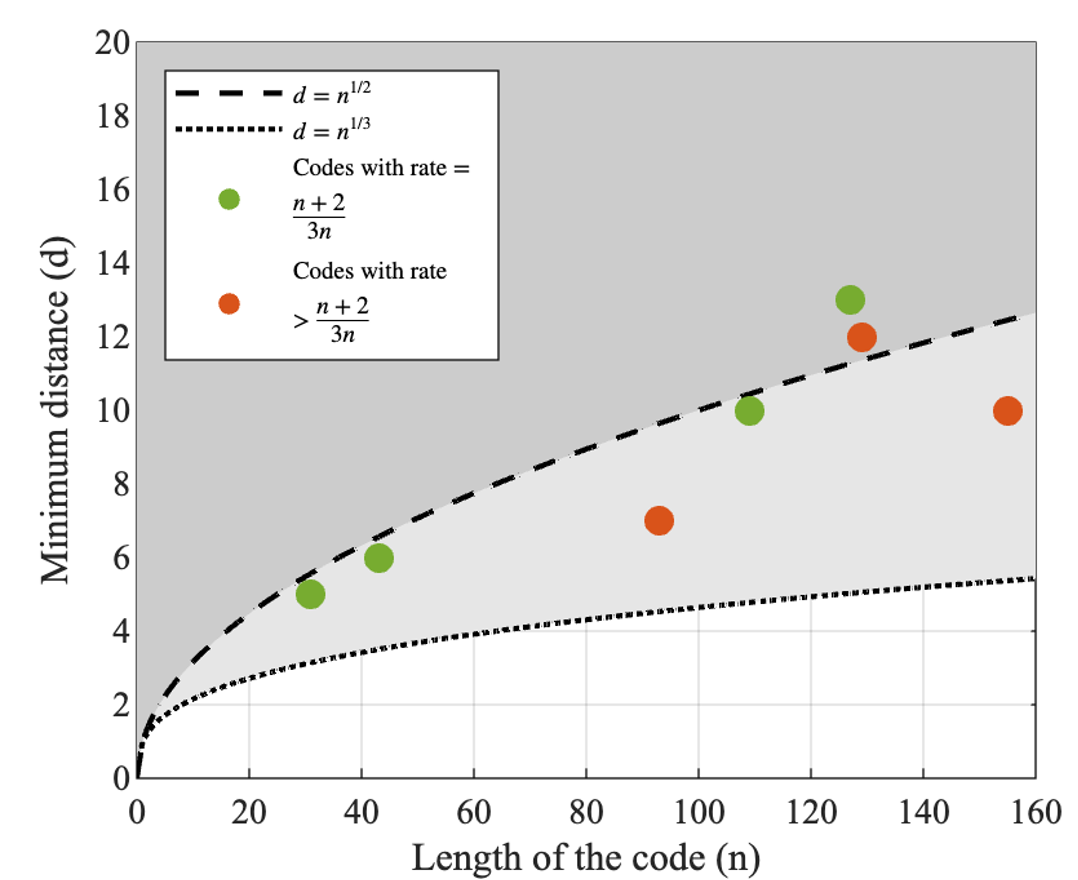}
\end{figure}

Figure~\ref{f:quantum_codes_graph} also illustrates the minimum distance of quantum triadic codes relative to the code's length. 
The diagram indicates that all minimum distances exceed the cubic root of the code's length. 
We anticipate similar behavior for quantum triadic codes of greater lengths.
Additionally, the codes depicted in black on the diagram exhibit a higher rate compared to the blue codes.

We have also made the Magma codes related to the splittings and constructions of cyclic, duadic, triadic, and related quantum codes available online at \cite{RezaGit}.

%%%%%%%%%%%%%%%%%%%%%%%%%%%%%%%%%%%%%%%%%%%%%%%%%%%%%%%%%%%%%%%%%%
\section{Conclusion}\label{S:Conc}
%%%%%%%%%%%%%%%%%%%%%%%%%%%%%%%%%%%%%%%%%%%%%%%%%%%%%%%%%%%%%%%%%%
To sum up, this study explored the construction of quantum CSS duadic codes with dimensions larger than one. 
We discussed a method for extending smaller splittings of quantum duadic codes to create larger, (degenerate) quantum duadic codes. 
Moreover, we introduced a technique for computing or bounding the minimum distances of duadic codes and their corresponding quantum codes constructed through the extended splittings. 
Additionally, we introduced quantum CSS triadic codes, a family capable of generating quantum codes with a rate of at least $\frac{1}{3}$. 

%We are currently analyzing the decoding performance of the quantum duadic and triadic codes discussed in this paper. The outcomes of this analysis will be incorporated into future versions of the paper.

\begin{credits}

%%%%%%%%%%%%%%%%%%%%%%%%%%%%%%%%%%%%
\subsubsection{\ackname}
%%%%%%%%%%%%%%%%%%%%%%%%%%%%%%%%%%%%
 The authors acknowledge support by the Spanish Ministry of Economy and Competitiveness through the MADDIE project (Grant No. PID2022-137099NB-C44), by the Spanish Ministry of Science and Innovation through the project “Few-qubit quantum hardware, algorithms and codes, on photonic and solid-state systems” (PLEC2021-008251), by the Ministry of Economic Affairs and Digital Transformation of the Spanish Government through the QUANTUM ENIA project call - Quantum Spain project, and by the European Union through the Recovery, Transformation and Resilience Plan - NextGenerationEU within the framework of the “Digital Spain 2026 Agenda”.
\subsubsection{\discintname}
The authors have no competing interests to declare that are
relevant to the content of this article. 
\end{credits}

\bibliographystyle{splncs04}
\bibliography{MyReferences}

\end{document}